\documentclass[prb,aps,amsfonts,showpacs,longbibliography,twocolumn,notitlepage,groupedaddress,floatfix,10pt,nofootinbib]{revtex4-2}
\usepackage{graphicx}
\usepackage[dvipsnames]{xcolor}
\usepackage{amsmath}
\usepackage{amssymb}
\usepackage{microtype}
\usepackage{verbatim}
\usepackage{placeins}
\usepackage[colorlinks=true, urlcolor=violet, linkcolor=blue, citecolor=blue, hyperindex=true, linktocpage=true]{hyperref}

\allowdisplaybreaks

\usepackage{physics}

\newcommand{\ii}{\mathrm{i}}
\newcommand{\ee}{\mathrm{e}}
\newcommand{\gammamr}{\gamma_{\mathrm{mr}}}

\newcommand{\gammaee}{\gamma_{\mathrm{ee}}}
\newcommand{\tauee}{\tau_{\mathrm{ee}}}
\newcommand{\pscatter}{p_{\mathrm{scatter}}}

\newcommand{\pd}{\partial}

\begin{document}

\title{Characterizing electronic scattering rates with transport in multiterminal devices}
\author{Jack H. Farrell}
\email[Contact author: ]{jack.farrell@colorado.edu}
\affiliation{Department of Physics and Center for Theory of Quantum Matter, University of Colorado Boulder, Boulder, Colorado 80309, USA}
\author{Andrew Lucas}
\affiliation{Department of Physics and Center for Theory of Quantum Matter, University of Colorado Boulder, Boulder, Colorado 80309, USA}

\begin{abstract}
Strongly interacting electrons in clean two-dimensional devices are theorized to exhibit many distinct transport regimes, such as ballistic, hydrodynamic, or diffusive.  Realistic samples often lie in crossover regimes between these idealized limits. We show how a single experiment on a multiterminal device can distinguish these regimes and constrain the relevant scattering rates without space-resolved imaging. Using a linearized Boltzmann model in a five-terminal geometry, we find that current partition among the drain contacts diagnoses the ballistic-hydrodynamic-ohmic crossover and allows extraction of momentum-relaxing and momentum-conserving scattering rates in the crossover regime. The same geometry also exhibits clear signatures of the tomographic regime, potentially allowing for a quantitative discrimination between viscous and tomographic flow in experiments. Our results demonstrate that multiterminal devices are a simpler experimental route to characterize transport regimes in electron liquids, relative to space-resolved imaging experiments.
\end{abstract}

\maketitle

\section{Introduction}
\begin{figure}[t]
    \centering

    \includegraphics[width=\columnwidth]{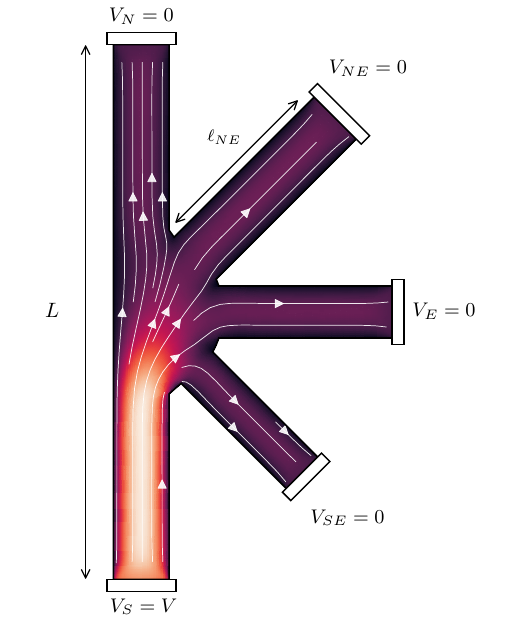}
    \caption{Schematic of the five-terminal ``fan'' geometry studied in this work.
    The color map represents the calculated current magnitude $|\vb{j}(\vb x)|$ obtained from the linearized Boltzmann equation \eqref{eq:boltzmann} with $\gamma_{\mathrm{mr}}=0$ and $\gamma_{\mathrm{ee}}=\gamma_3=100\, v_F / L$ in \eqref{eq:collisions}. The drain arms are chosen to have different widths and lengths, while $w/\ell$ is held fixed for each arm so that they represent identical ohmic channels. The relative arm dimensions are $\ell_N/L=\ell_S/L=0.35$, $\ell_{NE}/L=0.37$, $\ell_E/L=0.32$, and $\ell_{SE}/L=0.28$, with $w_a/\ell_a=0.30$ for all arms.  The relative angles are $45^\circ$.}
\label{fig:schematic}
\end{figure}
Hydrodynamics is the effective long-wavelength theory describing how conserved quantities evolve once rapid collisions establish local equilibrium~\cite{landauFluidMechanics2012, gurzhiHydrodynamicEffectsSolids1968,lucasHydrodynamicsElectronsGraphene2018, fritzHydrodynamicElectronicTransport2024, huiHydrodynamicsElectronicFermi2025}. In this regime, electron transport is not organized by the full nonequilibrium distribution function, but by a small set of slowly varying fields and transport coefficients, making it possible to describe experiments with relatively few phenomenological parameters. From a theoretical perspective, electronic hydrodynamics offers a controlled setting in which to connect microscopic scattering to emergent many-body behavior and to probe possible universal limits on transport in metals~\cite{Hartnoll, lucasResistivityBoundHydrodynamic2017, qiPlanckianBoundLocal2026}. From a technological perspective, the characteristic frequencies of micron-scale mesoscopic devices naturally fall in the high-gigahertz to low-terahertz window, opening the possibility of compact circuit elements built from viscous, inertial, or compressible flow~\cite{dyakonovShallowWaterAnalogy1993, mendlDyakonovShurInstabilityBallistictohydrodynamic2018,farrellTerahertzRadiationDyakonovShur2022d, geursSupersonicFlowHydraulic2025a, liongSpontaneousRunningWaves2025}.

At the same time, experiments in two-dimensional conductors do not probe the hydrodynamic regime in isolation, but rather access a sequence of qualitatively distinct regimes including nearly collisionless semiclassical ballistic flow~\cite{beenakkerQuantumTransportSemiconductor1991}, intermediate tomographic dynamics with a large number of nearly conserved quantities~\cite{ledwithTomographicDynamicsScaleDependent2019, hofmannCollectiveModesInteracting2022, nilssonNonequilibriumRelaxationOddEven2025}, viscous hydrodynamics, and ordinary diffusive (ohmic) transport~\cite{ashcroftSolidStatePhysics1976, girvinModernCondensedMatter2019}. While bulk observables such as nonlocal conductivity~\cite{levitovElectronViscosityCurrent2016,bandurinNegativeLocalResistance2016a, bandurinFluidityOnsetGraphene2018a}, superballistic conductance~\cite{guoHigherthanballisticConductionViscous2017, sternHowElectronHydrodynamics2022, kumarImagingHydrodynamicElectrons2022}, and the Gurzhi effect~\cite{gurzhiHydrodynamicEffectsSolids1968, dejongHydrodynamicElectronFlow1995, mollEvidenceHydrodynamicElectron2016} can qualitatively delineate pairs of these regimes such as hydrodynamic--diffusive or ballistic--hydrodynamic, sharply distinguishing all of them---and in particular the intermediate tomographic regime---is challenging, though magnetoresistance in annular devices and high order cyclotron resonance each appear promising~\cite{zengQuantitativeMeasurementViscosity2024, moiseenkoTestingTomographicFermi2025}.  Indeed, quantitatively determining microscopic scattering parameters has historically required sophisticated spatial imaging to resolve phenomena including Poiseuille channel flow profiles~\cite{sulpizioVisualizingPoiseuilleFlow2019a,zhangImagingFlatBand2026}, viscous backflow~\cite{levitovElectronViscosityCurrent2016, bandurinNegativeLocalResistance2016a}, or vorticity~\cite{aharon-steinbergDirectObservationVortices2022,zhangImagingFlatBand2026}.

In this work, we argue that full space resolution is not required to quantitatively measure microscopic scattering parameters in an electron fluid; one only requires multiterminal devices with enough contacts or voltage probes.  Our proposed geometry in Fig.~\ref{fig:schematic} operates on the simple principle that current partition between multiple drains is strongly regime dependent.
Ballistic carriers preferentially follow nearly straight trajectories into aligned drains, whereas hydrodynamic and diffusive currents readily redistribute around bends and obstacles. Hydrodynamic and diffusive transport can in turn be distinguished by their different dependence on channel width and geometric constrictions. By exploiting these distinct geometric responses, one can construct bulk observables that, if measured with high precision, remain directly sensitive to the underlying scattering mechanisms. Although for convenience we have focused on a single geometry in this paper, we stress that the general strategy is not sensitive to details of this geometry.

To quantitatively relate multiterminal observables to microscopic transport parameters, we present an open-source numerical framework \cite{farrell_fermisea_2026} for efficiently solving the linearized Boltzmann transport equation for isotropic Fermi liquids near $T=0$ in arbitrary two-dimensional (2D) domains with any number of contacts and a variety of realistic boundary conditions. These methods have recently been used to quantitatively model both space-resolved~\cite{zhangImagingFlatBand2026} and bulk observables~\cite{holleisCryogenicShockExfoliation2026} in dc settings. Our phenomenological scattering model features three parameters, namely $\gammamr$ (a momentum-relaxing scattering rate due to impurities and phonons), $\gammaee$ (the momentum-conserving electron-electron scattering rate), and $\gamma_3$ (which parametrizes the even-odd structure inherited from the microscopic Fermi-golden-rule collision kernel), each defined in \eqref{eq:collisions}.

We find that the three independent current fractions through the $NE$, $E$, and $SE$ arms of Fig.~\ref{fig:schematic} define a well-conditioned inverse problem for $\gammaee$ and $\gammamr$. We further show that multiterminal observables within the same framework are also sensitive to tomographic transport, allowing one to identify the onset of the intermediate regime and to place quantitative constraints on $\gamma_3$.  These results offer a quantitative way to distinguish the tomographic regime from viscous or ballistic transport in present-day experiments (where scaling predictions \cite{ledwithTomographicDynamicsScaleDependent2019, ben-shacharMagnetotransportTomographicElectrons2025} are not easy to detect using only a small range of sample sizes).

\section{Methods}
\begin{figure*}[t]
    \centering
    \includegraphics[width=\textwidth]{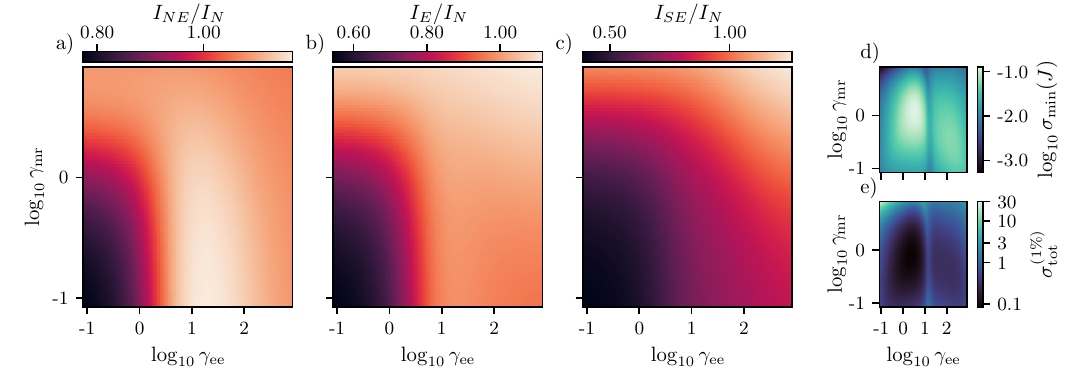}
    \caption{Current partition in the five-terminal fan geometry as a function of momentum-relaxing and momentum-conserving scattering. (a)--(c) Current $I_a$ for $a=NE, E, SE$, as a fraction of $I_N$, the current through the north contact, $I_N$. (d) Smallest singular value $\sigma_{\min}(J)$ of the Jacobian $J=\partial(I_{NE}/I_N,I_E/I_N,I_{SE}/I_N)/\partial(\log_{10}\gamma_{\mathrm{mr}},\log_{10}\gamma_{\mathrm{ee}})$, measuring the weakest local sensitivity of the observables to the scattering rates. (e) Propagated total $1\sigma$ statistical uncertainty $\sigma_{\mathrm{tot}}$, combining the two logarithmic-rate standard errors in quadrature, in decades, in the inferred scattering rates assuming independent $1\%$ relative standard errors in the four raw drain currents, as described in Appendix~\ref{app:inference}. These uncertainties are conditioned on the forward model and exclude systematic errors from uncertain geometry, contact properties, and other model inaccuracies. The crossover regime is the most favorable for extracting both $\gamma_{\mathrm{mr}}$ and $\gamma_{\mathrm{ee}}$ from multiterminal current measurements, while the strongly ohmic regime becomes ill conditioned.}\label{fig:fractions}
\end{figure*}

We study fermionic quasiparticles in a single isotropic band with dispersion $\varepsilon(|\vb p|)$.  With $\theta \equiv \arctan (p_y / p_x)$, in the limit $T \rightarrow 0$, dynamics are confined to the Fermi surface and we may take the deviation from equilibrium to be independent of $p \equiv |\vb p|$ according to
\begin{align}
    \delta f(\vb x, \vb p) = \phi(\vb x, \theta)\, \delta(\mu_0 - \varepsilon(p)).
\end{align}
Integrating the Boltzmann transport equation for $f$ over the radial momentum coordinate $p$ leads to a linearized Boltzmann equation for $\phi(\vb x, \theta)$, namely
\begin{align}\label{eq:boltzmann}
    \pd_t\phi + \vb{v}_F\cdot \nabla \phi = W[\phi],
\end{align}
where $\vb{v}_F = v_F\, \vu p$ is the Fermi velocity, $W[\phi]$ is the linearized fermionic collision integral, and we have set external forcing $\vb F = 0$, biasing flow instead through boundary conditions on $a_0$.  In an isotropic Fermi liquid, several recent studies~\cite{hofmannAnomalouslyLongLifetimes2023, thuillierMultipolarFermiSurface2025, thuillierACFingerprints2D2026} have shown that the linearized collision integral $W[\phi]$ is well approximated by
\begin{align}\label{eq:W}
W[\phi(\theta)]=-
\int_0^{2\pi}\frac{\dd{\theta'}}{2\pi}\,
\sum_{m=-\infty}^{\infty}\gamma_{m}\,\ee^{\ii m(\theta-\theta')}\phi(\theta'),
\end{align}
with relaxation rates $\gamma_m$ of respective Fourier components of $\phi$ given by
\begin{subequations}\label{eq:collisions}
\begin{align}
    \gamma_0 &= 0,\\
    \gamma_{\pm1} &= \gammamr,\\
    \gamma_m &= \left\lbrace \begin{array}{ll} \gammamr + \gammaee  &\ |m|\ge 2\text{ even} \\ \gammamr + \min\qty(\gammaee, \gamma_3 \qty(\frac{|m|}{3})^4) &\ |m|\ge 2\text{ odd} \end{array}\right..
\end{align}
\end{subequations}
The meaning of \eqref{eq:W} is that in a multipolar expansion of the Fermi surface $\phi(\theta) = \sum_{m} a_m \mathrm{e}^{\ii m \theta}$, different modes relax at different rates determined from Fermi's golden-rule expression for electron--electron collisions at low temperature. The three control parameters for $W[\phi]$ are $\gammamr$ (momentum-relaxing scattering from short-range impurities or phonons), $\gammaee$ (short-range electron-electron scattering), and $\gamma_3$ which sets the scale of the relaxation of the odd modes in two-body electron-electron collisions (which itself is controlled by $T/T_F$ or the thickness of the smeared Fermi surface at finite temperature).

We consider 2D devices with multiple metallic contacts $a \in \{1,\ldots,N\}$.

In linear response, the relation between the currents $I_a$ flowing into external contacts and the electrochemical potentials $V_b$ imposed at those contacts is
\begin{align}
    I_a = \sum_{b=1}^{N} G_{ab} V_b,
\end{align}
where $G_{ab}$ is a conductance matrix relating the potential imposed at contact $b$ to the current measured at contact $a$.  Importantly, $G_{ab}$ does not have $N^2$ independent entries.  For example, current conservation and gauge invariance imply
\begin{align}
    \sum_{a=1}^{N} I_a = 0,
    \qquad
    \sum_{a=1}^{N} G_{ab} = 0,
    \qquad
    \sum_{b=1}^{N} G_{ab} = 0,
\end{align}
so that the physically meaningful transport data furnish an $(N-1)$-dimensional subspace of voltages and currents.  Additional permutation symmetries of the device may further reduce the independent components of $G_{ab}$. In principle, one could fully characterize a given geometry by reconstructing the entire reduced conductance matrix, giving a complete bulk linear-response description of the device. Measurements of $I_a$ could be performed using separate preamplifiers~\cite{saminadayarObservation$mathitemathit3$1997}, in-series small resistors~\cite{dattaElectronicTransportMesoscopic2009}, Hall effect measurements in the presence of a small magnetic field~\cite{geimVanWaalsHeterostructures2013}, scanning tunneling microscopy~\cite{binnigSurfaceStudiesScanning1982}, or scanning magnetometry~\cite{sulpizioVisualizingPoiseuilleFlow2019a}.  The advantage of multiterminal experiments compared to comparing different devices instead is that the latter is susceptible to fabrication discrepancies between samples while the former allows different regimes to be characterized on a single device stack.\footnote{Spatially-localized impurities in the micron-scale device can complicate the quantitative comparison between theory and experiment whether or not one makes one multiterminal device or multiple scaled copies of the same device.}

Still, we are sensitive to the experimental challenges of quantitatively holding multiple contacts at known potentials relative to a virtual ground. Accordingly, in this paper, we focus on a particularly simple protocol in which one contact (the ``source,'' $a = S$) is held at a potential $V_S = V$ while the remaining contacts (``drains'') $a \neq S$ are grounded at $V_a = 0$. Branching ratios---the fraction of current entering each contact---offer simple bulk observables that cancel an overall multiplicative current scale. This cancellation does not remove sensitivity to contact nonidealities that change the current partition, such as branch-dependent contact resistance or imperfect carrier absorption. Quantitative inference therefore remains conditional on the assumed contact model.  We have performed additional numerical simulations for  floating voltage probes rather than grounded contacts, which may be a more familiar experiment than current branching---see Appendix~\ref{sec:contacts}. In the main text, we focus on the grounded-contact experiment.

For boundary conditions, we model walls as partially diffuse and partially specular---see Appendix~\ref{app:numerics}.  We denote the probability of diffuse wall scattering by $\pscatter$, so that $\pscatter=0$ is fully specular and $\pscatter=1$ is fully diffuse.  For simplicity, the main-text results use $\pscatter=1$, as we expect that intentionally fabricating devices with microscopically coarse walls, while counterintuitive, is experimentally tractable. Appendix~\ref{sec:bcsensitivity} quantifies the error incurred when data with $\pscatter<1$ are fitted using the fully diffuse model.  The parameter $\pscatter$ could instead be included as an additional unknown at the cost of complicating the fit.  If a three-parameter fit is not desirable, $\pscatter$ may in general be estimated independently from ballistic reflection amplitudes or magnetic focusing~\cite{leeBallisticMinibandConduction2016,bachmannSupergeometricElectronFocusing2019, mcguinnessLowsymmetryNonlocalTransport2021a,moravecDirectionalBallisticMagnetotransport2026,holleisCryogenicShockExfoliation2026}, though we acknowledge that this requires an additional experiment.

\section{Results}
For concreteness, in this work, we consider the five-terminal ``fan'' sketched in Fig.~\ref{fig:schematic}. Two design principles motivated our choice of geometry.  First, in the ballistic limit, current should preferentially flow into the branch labeled $N$. Secondly, the three branches labeled $SE, E$, and $NE$ differ in length and width in such a way that they represent identical ohmic channels: recall that, if the device has homogeneous resistivity $\rho$, then the effective resistance of each channel is given by
\begin{equation}\label{eq:ohmic}
    R_{\mathrm{ohmic}} = \rho \frac{\ell_a}{w_a}.
\end{equation}
Since we keep $\ell_a/w_a$ fixed for each channel, the ohmic resistance would therefore be identical in the idealized limit where each $\ell_a \rightarrow \infty$. On the other hand, in the hydrodynamic limit $\gammaee \gg 1$ (in units of $v_F/L$), Poiseuille flow predicts a stronger scaling with $w_a$~\cite{gurzhiHydrodynamicEffectsSolids1968,lucasHydrodynamicsElectronsGraphene2018, fritzHydrodynamicElectronicTransport2024, huiHydrodynamicsElectronicFermi2025},
\begin{align}\label{eq:hydro}
R_{\rm hydro}
\sim
\gammaee^{-1} \frac{3\,p_F v_F}{n e^2}
\frac{\ell_a}{w_a^3},
\end{align}
so we expect the current fractions through branches labeled $NE$, $E$, and $SE$ to differentiate ohmic and hydrodynamic current flow.

\subsection{Ballistic-hydrodynamic crossover}
To begin, we restrict our attention to $\gammaee$ and $\gammamr$ (taking $\gammaee=\gamma_3$; ignoring tomographic effects), which is enough to quantitatively resolve the ballistic, hydrodynamic, and diffusive regimes.  Indeed we have already applied the numerical methods introduced in this work to determine $\gammaee$ and $\gammamr$ in a recent space-resolved imaging experiment~\cite{zhangImagingFlatBand2026}.  Here we show that a similar determination could be performed via multiterminal transport experiments instead.

We have solved the linear Boltzmann equation \eqref{eq:boltzmann} in the geometry and biasing configuration of Fig.~\ref{fig:schematic}, assuming diffuse walls (Appendix~\ref{app:numerics}). After performing the space-resolved simulations, we measure the current $I_a$ through each contact and plot the fractions $I_a/I_N$. Figures~\ref{fig:fractions}(a)--\ref{fig:fractions}(c) show heatmaps of these fractions as functions of $\gammaee$ and $\gammamr$.

The simulation results confirm our qualitative expectations for the behavior of the current fractions across the three regimes of interest. For the $NE$, $E$, and especially $SE$ branches, $I_{NE}/I_N$, $I_E/I_N$, and $I_{SE}/I_N$ show suppression in the ballistic limit (lower-left corner); ballistic carriers preferentially follow straight-line trajectories. On a qualitative level, $I_a / I_N < 1$ is an extremely clear signature of ballistic flow, and indeed we find $I_{SE} / I_{N} \sim 0.5$ for the backward-facing branch in the ballistic limit. On the other hand, as $\gammaee$ increases at low $\gammamr$, the current through $NE$ noticeably exceeds that through $SE$, in agreement with \eqref{eq:hydro}, given that the branch labeled $SE$ is narrower than $NE$.  As such, $I_{NE}/I_{SE} > 1$, with $I_{NE} / I_N \geq 1$ ruling out ballistic flow, offers an unmistakable qualitative diagnostic of viscous-dominated flow.  While the quantitative predictions of Eqs.~\eqref{eq:ohmic} and \eqref{eq:hydro} do not hold in our proposed device due to strong finite-size effects ($\ell_a/w_a$ is not large), the device succeeds in differentiating ohmic, hydrodynamic, and ballistic transport.

Indeed, beyond qualitatively distinguishing transport regimes, the three current fractions in Figs.~\ref{fig:fractions}(a)--\ref{fig:fractions}(c) display nondegenerate dependence on $\gammaee$, $\gammamr$ in the crossover regime, so we expect these three observables may be used to quantitatively ``triangulate'' the scattering rates. To characterize the sensitivity to $\gammaee$ and $\gammamr$, we consider the Jacobian $J=\partial(I_{NE}/I_N,I_E/I_N,I_{SE}/I_N)/\partial(\log_{10}\gamma_{\mathrm{mr}},\log_{10}\gamma_{\mathrm{ee}})$. Its minimum singular value $\sigma_{\text{min}}$ represents the weakest local sensitivity of the observables to $\gammaee$ and $\gammamr$ and it is displayed in Fig.~\ref{fig:fractions}(d). The crossover regime $\gammaee \sim 1\, v_F/L$ is particularly promising for quantitative estimation of both phenomenological parameters.

To estimate how precisely a single measurement can determine the scattering rates, Fig.~\ref{fig:fractions}(e) shows the total propagated $1\sigma$ uncertainty in the two logarithmic rates, assuming independent $1\%$ relative standard errors in the four raw drain currents as described in Appendix~\ref{app:inference}. These statistical uncertainties assume that the kinetic model is correct and exclude systematic errors from uncertain geometry, contact properties, wall scattering, or other model inaccuracies; Appendix~\ref{sec:bcsensitivity} separately illustrates the bias caused by an incorrect wall model. Qualitatively, the standard error at fixed relative error in $I_a$ follows the local sensitivity represented in panel (d), indicating that the singular value analysis is appropriate. In the region of poor conditioning, a $1 \%$ uncertainty in the raw currents gives formal local standard errors up to approximately $15$ decades, signaling that the local inversion no longer provides a useful quantitative constraint.  Nonetheless, the well-conditioned crossover region offers precision as high as $0.09$ decades, even with $1 \%$ error in the raw currents, from a \emph{single-shot} measurement with exactly \emph{one} biasing configuration. We note that more sophisticated biasing arrangements, especially if they were continuously varied, would in principle further reduce uncertainty and permit determination of arbitrary numbers of phenomenological parameters such as $\gamma_3$ and even $\pscatter$, but to exhaustively characterize the large space of possible experiments realizing our design principles is beyond the scope of this work. Still, in the next subsection, we study one plausible explicit experimental setup achieving a high-precision determination of $\gammamr$ and $\gammaee$.

\subsection{Mock experiment}\label{sec:mock-experiment}
A realistic experiment may sweep $\gammaee$ and $\gammamr$ together
as functions of a control parameter such as temperature.  The local
sensitivity of the map
$(I_{NE}/I_N,I_E/I_N,I_{SE}/I_N)\mapsto
(\gammaee,\gammamr)$ therefore provides a conservative estimate of the
information available from a full temperature sweep.  As an end-to-end
test, we consider clean encapsulated Bernal bilayer graphene at carrier
density $n_0=1.5\times10^{12}\,\mathrm{cm}^{-2}$ and choose the device
scale so that the source width is $1\,\mu\mathrm{m}$.  Experiments and first-principles calculations~\cite{bandurinFluidityOnsetGraphene2018a,borysenkoElectronPhononInteractions2011} appear broadly consistent with the simple model
\begin{subequations}\label{eq:mock-rate-laws}
\begin{align}
    \gammaee(T) &=
    18.153\left(\frac{T}{100\,\mathrm{K}}\right)^2\frac{v_F}{L},\\
    \gammamr(T) &=
    \left[0.667+
    2.119\left(\frac{T}{100\,\mathrm{K}}\right)\right]\frac{v_F}{L}.
\end{align}
\end{subequations}
We use this model to generate solutions of the Boltzmann equation at $14$ temperatures from $20$ to $150\,\mathrm{K}$ in $10\,\mathrm{K}$ steps, taking $\gamma_3=\gammaee$ throughout. Then we add independent $1\%$ Gaussian multiplicative noise to each of the four drain currents at each temperature and fit all temperatures jointly for $10^4$ noise realizations, leaving the three coefficients in the assumed temperature dependence free. We fit the logarithms of the current ratios, including their correlations due to the shared $I_N$ measurement. Appendix~\ref{app:inference} gives the fitting objective and constraints.

\begin{figure}[t]
    \centering
    \includegraphics[width=\columnwidth]{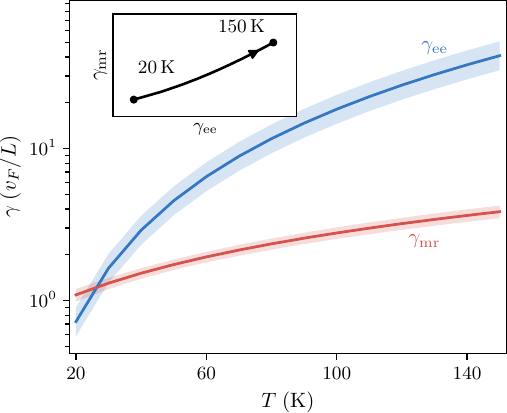}
    \caption{Mock experiment for the temperature-dependent momentum-conserving and momentum-relaxing rates. The solid blue and red traces show the median recovered $\gammaee(T)$ and $\gammamr(T)$ respectively, and the shaded bands indicate the central $95\%$ pointwise statistical intervals across noise realizations, assuming that the kinetic model and temperature dependences are correct. Systematic errors from uncertain geometry, contact properties, and other model inaccuracies are excluded.  The inset shows the input bilayer-graphene trajectory in $(\gamma_{\mathrm{ee}},\gamma_{\mathrm{mr}})$ space, with the arrow indicating increasing temperature. The reconstruction uses $10^4$ realizations with independent $1\%$ Gaussian multiplicative noise on each drain current, sampled at $14$ temperatures from $20$ to $150\,\mathrm{K}$ in $10\,\mathrm{K}$ steps and fitted jointly as described in Appendix~\ref{app:inference}. Across $20$--$150\,\mathrm{K}$, the central $95\%$ errors remain within $-0.095$ to $+0.094$ decades for $\gammaee$ and $-0.045$ to $+0.040$ decades for $\gammamr$.}
\label{fig:mock-experiment}
\end{figure}

Figure~\ref{fig:mock-experiment} reports the resulting uncertainty directly in the rates. The central $95\%$ intervals remain within $0.10$ decades of the input $\gammaee$ law and within $0.05$ decades of the input $\gammamr$ law over the full temperature range. Again, these pointwise statistical intervals do not include systematic errors from uncertain geometry, contact properties, wall scattering, or incorrect temperature dependences.

\begin{figure*}[t]
    \centering
    \includegraphics[width=\textwidth]{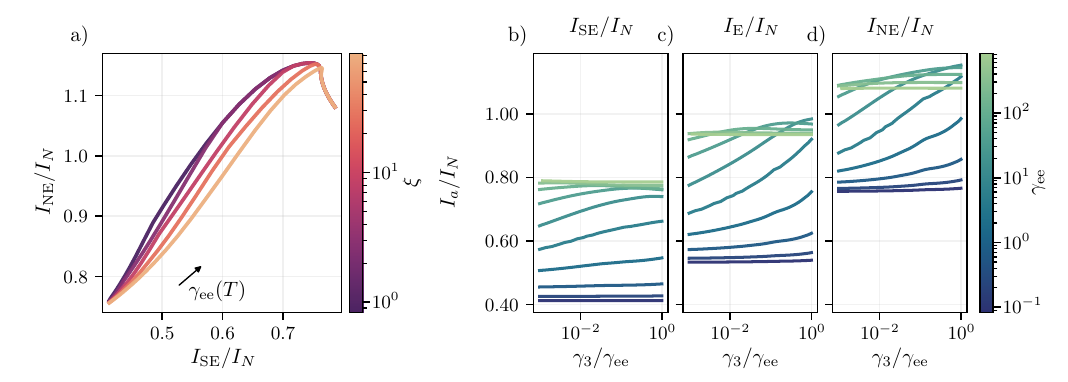}
   \caption{Emergence of tomographic physics. We set $\gammamr = 0$ and consider the fraction $I_{a} / I_{N}$ for $a = SE, E, NE$.  (a) Sample trajectories in phase space assuming $\gammaee\sim T^2$ and $\gamma_3 \sim T^4$, so that $\gamma_3 = \gammaee^2 / \xi$, with $\xi$ controlling the strength of even-odd effects.  At large $\xi$ (tomographic regime), the trajectory in $(I_{SE}/I_N, I_{NE}/I_N)$ space develops a sharp feature as $T$ is swept. (b)--(d) $I_a/I_N$ as a function of $\gamma_3 / \gammaee$, artificially treating them as independent parameters, expressing a rich tomographic enhancement or suppression of current branching ratios.}\label{fig:tomographic}
\end{figure*}

\subsection{Tomographic flow}\label{sec:tomo}
Multiterminal observables are also sensitive to the details of short-range fermion-fermion scattering in 2D modeled by $\gamma_3 < \gammaee$ in \eqref{eq:collisions}. In an experiment, the emergence of this so-called ``tomographic'' flow will most naturally be accessed by varying temperature $T$, which we may model by assuming forms for $\gammaee(T)$, $\gamma_3(T)$.  At low temperature, fermion-fermion scattering is dominated by nearly head-on collisions, which leads to $\gammaee\sim (T/T_F)^2$ [up to $\ln(T/T_F)$ corrections]~\cite{girvinModernCondensedMatter2019}. By inversion symmetry, the nearly head-on scattering cannot relax the inversion-odd component of the distribution function and the relaxation rates of odd harmonics are found to be suppressed as $\gamma_3\sim (T/T_F)^4$~\cite{ledwithTomographicDynamicsScaleDependent2019}.  We therefore parametrize $\gamma_3 = \gammaee^2/\xi$, where $\xi$ is an inverse time-scale measuring the strength of odd-even effects. Figure~\ref{fig:tomographic}(a) shows the phase trajectory for $I_{SE}/I_{N}$ and $I_{NE}/I_{N}$ as $\gammaee$ ($T^2$) varies.  In the ballistic limit $\gammaee \rightarrow 0$ and the hydrodynamic limit $\gammaee \rightarrow \infty$, the trajectories coincide. When $\xi$ is large, the trajectory develops a sharp corner at intermediate values of $\gammaee$ and follows a distinct curve in $(I_{SE}$, $I_{NE})$ space. Current branching ratios thus offer sharp signatures of the tomographic regime accessible by experimental control parameters.

Further, assuming that $\gammaee$ and $\gamma_3$ are independent,  in Figs.~\ref{fig:tomographic}(b)--\ref{fig:tomographic}(d), we demonstrate the consequences of odd-even effects on transport through the fractions $I_{a}/I_{N}$ as functions of $\gamma_3$ at fixed $\gammaee$. For all three, $\gamma_3 < \gammaee$ suppresses the fraction of currents through the side channels, in agreement with the expectation that tomographic flow allows increased directional memory relative to the hydrodynamic regime at higher temperatures. In particular, $(I_{NE} / I_N - 1)$ shows a sign change as a function of $\gamma_3$ at intermediate $\gammaee$, so $I_{NE} / I_N$ may offer a qualitative experimental signature of the tomographic regime, which may be easier to interpret than subtle $T$ or $w$ dependence of conductances in a crossover regime.  It would be interesting to study whether our kinetic results are captured in low-order closures of the Boltzmann equation that have been proposed to model the tomographic regime~\cite{ben-shacharMagnetotransportTomographicElectrons2025,ben-shacharTomographicElectronFlow2025}.  Similarly, our multiterminal approach may be useful for calibrating high-order cyclotron resonance measurements to directly probe the $\gamma_m$ for higher multipolar deformations~\cite{moiseenkoTestingTomographicFermi2025}.  Our results provide a complementary experimental consequence of the tomographic dynamics studied in earlier work. The even-odd hierarchy has been connected to scale-dependent transport, collective modes, and modified boundary conditions in confined geometries~\cite{ledwithTomographicDynamicsScaleDependent2019,hofmannCollectiveModesInteracting2022,ben-shacharTomographicElectronFlow2025}, although the $m$ dependence of the odd rate (resulting in a Navier-Stokes equation with fractional powers of $\nabla$) is neglected in these works. Here, we show how this hierarchy affects current partition in a single multiterminal device. The effect is not small, as at $\gammaee\approx5\,v_F/L$ and $\gammamr=0$ reducing $\gamma_3/\gammaee$ from $1$ to $10^{-3}$ changes $I_{NE}/I_N$ from approximately $1.118$ to $0.877$, crossing unity. For comparison, annular magnetoresistance separates ohmic and viscous contributions and probes their temperature dependence~\cite{zengQuantitativeMeasurementViscosity2024}, while high-order cyclotron resonance probes individual angular-harmonic relaxation rates through resonance linewidths~\cite{moiseenkoTestingTomographicFermi2025}. Our proposed measurement accesses the same collision physics through zero-field dc transport, with quantitative interpretation supplied by the kinetic model $\gamma_3$.

\section{Conclusion}
We have shown that multiterminal transport in a single mesoscopic device provides a route for quantitatively diagnosing electronic transport regimes without the need for space-resolved imaging. In the five-terminal fan geometry studied here, simple current branching fractions already distinguish the ballistic, hydrodynamic, and ohmic limits, and in the crossover regime they define a well-conditioned inverse problem for the momentum-relaxing and momentum-conserving scattering rates. The same framework is also sensitive to the even--odd structure of electron-electron collisions, allowing tomographic transport to be identified through realistic experimental signatures. Our results illustrate that carefully designed device geometry can convert subtle information about the collision integral into robust bulk observables that are directly accessible in standard transport experiments.

One shortcoming of studying nonohmic transport physics in nonstandard 2D geometries is that it is often not obvious how to interpret data without detailed comparison to numerical models, which themselves often make many simplifying assumptions.  For example, as in this work, it is common to assume the Fermi surface is circular to simplify the numerical analysis, yet noncircular Fermi surfaces are known to have drastic effects on ballistic transport \cite{mcguinnessLowsymmetryNonlocalTransport2021a, okaBallisticTransportExperiment2019}.
It is, in principle, straightforward to extend our methods to more microscopic kinetic theories of transport which account for both realistic Fermi surface shapes as well as \emph{ab initio} collision integrals $W[\phi]$~\cite{coulterMicroscopicOriginsHydrodynamic2018,hofmannAnomalouslyLongLifetimes2023,thuillierMultipolarFermiSurface2025} and we plan to do this in future work.

It would be interesting to engineer optimal multiterminal geometries for measuring $\gammaee$, $\gammamr$, and $\gamma_3$, perhaps borrowing inverse-design techniques from the machine learning literature and minimizing an appropriate cost function.  Further, beyond the case of circular Fermi surfaces, we expect similar methods will permit the design of anisotropic viscometers motivated by the point-group symmetry of the underlying crystal lattice~\cite{cookViscometryElectronFluids2021}.

\section*{Acknowledgments}
We thank J. Geurs, D. Thuillier, X. Huang, C. Zhang, and E. Redekop for discussions.
This work was supported by the National Science Foundation via CAREER Grant No.~DMR-2145544.  This work utilized the Alpine high performance computing resource at the University of Colorado Boulder. Alpine is jointly funded by the University of Colorado Boulder, the University of Colorado Anschutz, Colorado State University, and the National Science Foundation (Award No.~2201538).

\section*{Data availability}
The data, Julia scripts to generate the data, and Python scripts to produce the figures are available~\cite{dataRepository}. The open source software used to simulate the Boltzmann equation can be found online~\cite{farrell_fermisea_2026}.

\appendix

\section{Voltage-biased floating-probe experiments}\label{sec:contacts}
In the same geometry as Fig.~\ref{fig:schematic}, we study an alternate biasing protocol that is closer to a conventional nonlocal-voltage measurement. The source contact $S$ is held at $V_S$, the east contact $E$ is grounded, and the north, northeast, and southeast contacts float. Their potentials are adjusted independently until $I_N=I_{NE}=I_{SE}=0$. The three measured observables are therefore the dimensionless voltage ratios $U_a\equiv V_a/V_S$ for $a=N,NE,SE$. Any permutation of voltage- and current-controlled branches is accessible to the same numerical method.

\begin{figure*}[t]
    \centering
    \includegraphics[width=\textwidth]{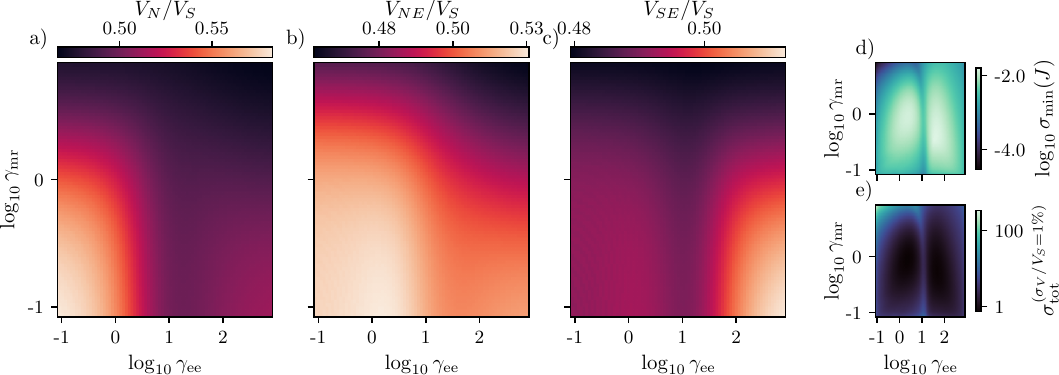}
    \caption{Voltage-biased floating-probe experiment in the fan geometry. The source $S$ is held at $V_S$, the east contact $E$ is grounded, and $N$, $NE$, and $SE$ float at zero net current. (a)--(c) The measured ratios $U_a=V_a/V_S$ for $a=N,NE,SE$. (d) Smallest singular value $\sigma_{\min}(J)$ of $J=\partial(U_N,U_{NE},U_{SE})/\partial(\log_{10}\gamma_{\mathrm{mr}},\log_{10}\gamma_{\mathrm{ee}})$, measuring the weakest local sensitivity to the two rates. (e) Total propagated $1\sigma$ uncertainty $\sigma_{\mathrm{tot}}=[\operatorname{tr}\operatorname{Cov}(\log_{10}\gamma_{\mathrm{mr}},\log_{10}\gamma_{\mathrm{ee}})]^{1/2}$, assuming an exact applied bias and independent additive voltage floors $\sigma_V/V_S=1\%$. The minimum and median values of $\sigma_{\mathrm{tot}}$ over the sweep are $0.74$ and $2.21$ decades, respectively.}\label{fig:fraction_contact}
\end{figure*}

Figure~\ref{fig:fraction_contact} shows the completed $79\times79$ sweep in $(\gammaee,\gammamr)$. To assess practical identifiability, we treat the imposed source bias as exact and assign each measured voltage ratio an independent additive $1\sigma$ floor $\sigma(U_a)=\sigma_V/V_S=0.01$. We then propagate these errors through the local Jacobian. The best-conditioned point has a total joint rate uncertainty of $0.74$ decades, while the grid median is $2.21$ decades. Thus this particular fan voltage-probe protocol retains sensitivity to both rates but, at the assumed voltage floor, is substantially less precise than the grounded-contact current-branching protocol in the main text.

\section{Sensitivity to Boundary Conditions}\label{sec:bcsensitivity}

\begin{figure*}[t]
    \centering
    \includegraphics[width=\textwidth]{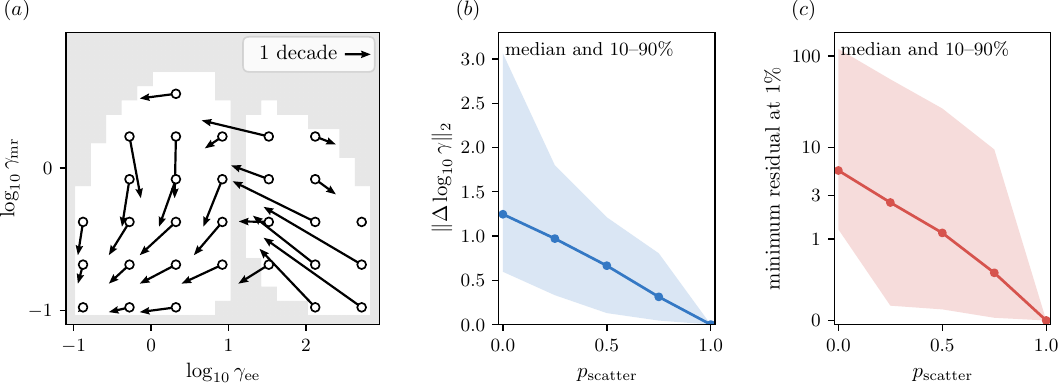}
    \caption{Sensitivity of the inferred scattering rates to diffuse wall scattering probability $p_\mathrm{scatter}$ as defined in \eqref{eq:maxwell_target_discrete}.  Synthetic branching ratios generated with diffuse-scattering probability $\pscatter$ are fitted using the fully diffuse model, $\pscatter=1$, to study the error.  (a) Displacement of the inferred rates for fully specular walls.  Circles mark the true values of $(\log_{10}\gammaee,\log_{10}\gammamr)$ and arrows point toward the inferred values. Gray background indicates points excluded by the uncertainty criterion described in the text.  (b) Median and $10$--$90\%$ range of the bias $\|\Delta\log_{10}\gamma\|_2$.  (c) Median and $10$--$90\%$ range of the minimum fit residual in units of the assumed $1\%$ current uncertainty.  Because three independent branching ratios constrain two rates, an incorrect wall model can leave a detectable residual rather than being absorbed entirely into the inferred rates.}\label{fig:wall-scattering}
\end{figure*}

To quantify sensitivity to the wall model, we generate synthetic current fractions over the same $(\gammaee,\gammamr)$ grid for $\pscatter=0$, $0.25$, $0.5$, and $0.75$, and fit them using the fully diffuse ($\pscatter=1$) map used in the main text.  We use the same three branching ratios, $I_a/I_N$ for $a=NE,E,SE$, in a weighted least-squares fit and assume $1\%$ relative uncertainty in each raw current.  We restrict the summary statistics in Fig.~\ref{fig:wall-scattering} to interior grid points where the fully diffuse model gives a total local $1\sigma$ uncertainty no larger than $0.5$ decades in the two inferred rates.

For $\pscatter=0.75$, the median rate bias is $0.31$ decades and the median residual is $0.59$ times the assumed measurement uncertainty.  A modest wall-model error can therefore bias the inferred rates without producing an obvious lack of fit.  The mismatch becomes clearer as the wall becomes more specular: at $\pscatter=0$, the median bias reaches $1.25$ decades and the median residual reaches $5.60$ times the assumed uncertainty.  The wall parameter should therefore be measured, controlled, or fitted jointly whenever the device is not known to be strongly diffuse.

\section{Linear Hydrodynamics}\label{app:hydrodynamics}
In this appendix, we briefly review the emergence of hydrodynamics from the linear Boltzmann equation in the limit $\gammamr \rightarrow 0$, $\gammaee \rightarrow \infty$ and make contact with the usual linearized Navier-Stokes treatment of electron hydrodynamics. A more detailed and general calculation may be found, for example, in Ref.~\cite{lucasHydrodynamicsElectronsGraphene2018}. For simplicity we take $\gamma_3 = \gammaee$ in what follows.  We will also assume a parabolic band, $\varepsilon = p^2 /2m$, so that the calculations may proceed explicitly, and we write $\tauee = 1/\gammaee$.
The hydrodynamic variables---number density $\delta n(\vb x, t) = n(\vb x, t) - n_0$ and momentum density $\vb*\pi(\vb x, t)$---are obtained by taking moments of the distribution function $f$. In particular,
\begin{subequations}\label{eq:moments}
    \begin{align}
        \delta n(\vb x, t) &= \frac{m}{2\pi\hbar^2}\int \frac{\dd \theta}{2\pi}\, \phi(\vb x, \theta, t),\\
        \vb*\pi(\vb x, t) &= \frac{m^2v_F}{2\pi\hbar^2} \int \frac{\dd \theta}{2\pi}\, \vu{p}(\theta) \phi(\vb x, \theta, t)\nonumber\\
        &= m n_0 \vb u(\vb x, t),
    \end{align}
\end{subequations}
where in the last equality we have introduced the fluid velocity $\vb u(\vb x, t)$ and used $n_0 = m^2 v_F^2 / 4\pi\hbar^2$.  We note that the coefficient in general differs from $m n_0$ for nonparabolic bands.
Next, from the explicit form of our approximate collision integral~\eqref{eq:collisions}, we may write the linear Boltzmann equation~\eqref{eq:boltzmann} explicitly as
\begin{equation}\label{eq:appboltzmann}
    \pd_t \phi + \vb{v}_F \cdot \nabla \phi =
    \frac{1}{\tauee}\qty(\frac{2\pi \hbar^2}{m}\delta n + m v_F \vb u \cdot \vu p - \phi).
\end{equation}
We will now solve \eqref{eq:appboltzmann} order by order in $\tauee$. When $\tauee \rightarrow 0$, the right-hand side of \eqref{eq:appboltzmann} can remain finite only if the bracket vanishes, so that $\phi = \phi^{(0)} =  \frac{2\pi \hbar^2}{m}\delta n + m v_F \vb u \cdot \vu p$.  In this so-called ideal hydrodynamic limit, taking moments of \eqref{eq:appboltzmann} and using \eqref{eq:moments}, we find the equations of motion (linearized Euler equations)
\begin{subequations}\label{eq:idealhydro}
\begin{align}
    \pd_t \delta n + n_0 \nabla \cdot \vb u &= 0,\\
    m n_0 \pd_t \vb u + \frac{mv_F^2}{2} \nabla \delta n &= 0.
\end{align}
\end{subequations}
The ideal hydrodynamic equations~\eqref{eq:idealhydro} describe sound waves propagating at speed $c_s = v_F / \sqrt{2}$.

Next, at first order in $\tauee$, we self-consistently write $\phi \approx \phi^{(0)} - \tauee \mathcal{P}\left( \pd_t \phi^{(0)} + \vb{v}_F \cdot \nabla \phi^{(0)}\right)$, where $\mathcal{P}$ projects onto angular harmonics $m \geq 2$; the $m = 0, 1$ components of the correction vanish because $\delta n$ and $\vb u$ are defined by the full moments~\eqref{eq:moments}.  We take $\tauee \pd_t \phi^{(0)} = 0$, the dc limit $\omega \tauee \ll 1$.   Substituting this self-consistent ansatz into \eqref{eq:appboltzmann} and taking moments, we find
\begin{subequations}\label{eq:viscoushydro}
    \begin{align}
    \pd_t \delta n + n_0 \nabla \cdot \vb u &= 0,\\
    m n_0 \pd_t \vb u + \frac{mv_F^2}{2} \nabla \delta n - \frac{m n_0 v_F^2\tauee}{4}\nabla^2 \vb u&= 0.\label{eq:stokes}
\end{align}
\end{subequations}
The latter equation \eqref{eq:stokes} is the linearized Navier-Stokes equation.  We identify $\eta = m n_0 v_F^2 \tauee/4$ as the dynamic shear viscosity, and the corresponding kinematic viscosity is $\nu \equiv \eta / (n_0 m) = v_F^2 \tauee/4$, in agreement with the usual result for the shear viscosity of 2D Fermi liquids~\cite{lucasHydrodynamicsElectronsGraphene2018}.  The corresponding bulk viscosity term vanishes, as it must for a Galilean-invariant parabolic band.

\section{Numerical methods}\label{app:numerics}
  \subsection{Harmonic truncation and streaming}
  Our goal is to numerically solve \eqref{eq:boltzmann}--\eqref{eq:collisions}. We truncate the
  Fourier expansion of $\phi(\vb{x},\theta,t)$ at harmonic order
  $M$, working in the real representation
  \begin{multline}
      \phi(\vb{x},\theta,t) \approx
      a_0(\vb{x},t)\\
      + \sum_{m=1}^{M}\bigl[
          a_m(\vb{x},t)\cos m\theta
        + b_m(\vb{x},t)\sin m\theta
        \bigr].
  \end{multline}
  This is equivalent to retaining $|m|\leq M$ in the complex expansion of the
  main text.  Note that we do \emph{not} carry the conventional factor of
  $\tfrac12$ on the monopole term; with this convention $a_0$ is recovered from
  $\phi$ with the angular weight $\tfrac1{2\pi}\!\int\!\dd\theta$, whereas every
  other harmonic uses $\tfrac1{\pi}\!\int\!\dd\theta$, which makes the $a_0$ row
  and column of the streaming matrices below asymmetric.  The local state is the
  vector $\vb{u} = (a_0,\, a_1,\, b_1,\,\ldots,\, a_M,\, b_M)^T \in
  \mathbb{R}^{1+2M}$.  The monopole $a_0 \propto \delta n$ tracks the charge
  density deviation, the dipole $(a_1, b_1)$ carries the current, and higher
  harmonics encode the angular anisotropy responsible for crossover physics
  between ballistic and hydrodynamic limits.

  Substituting the truncated expansion into~\eqref{eq:boltzmann} and applying the
  product identities such as $\cos\theta \cos m\theta = \tfrac{1}{2}[\cos(m{-}1)\theta +
  \cos(m{+}1)\theta]$, the free-streaming term $\vb{v}_F\cdot\nabla\phi$
  becomes the linear first-order system
  \begin{align}\label{eq:harmonic_system}
      \pd_t \vb{u} + A_x \pd_x \vb{u} + A_y \pd_y \vb{u} = C[\vb{u}],
  \end{align}
  where $A_x$ and $A_y$ are sparse $(1{+}2M)\times(1{+}2M)$ matrices and $C[\vb{u}]$
  collects the collision source terms.  The only nonzero entries of $A_x$ and
  $A_y$ are the monopole couplings
  \begin{subequations}
  \begin{align}
      (A_x)_{a_1,\, a_0} &= v_F, &
      (A_x)_{a_0,\, a_1} &= \tfrac{v_F}{2}, \\
      (A_y)_{b_1,\, a_0} &= v_F, &
      (A_y)_{a_0,\, b_1} &= \tfrac{v_F}{2},
  \end{align}
  together with the bulk couplings, valid for $m\geq1$ with the understanding
  that any neighboring index $m\pm1$ must itself be $\geq1$,
  \begin{align}
      (A_x)_{a_m,\, a_{m\pm 1}} &= \tfrac{v_F}{2}, &
      (A_x)_{b_m,\, b_{m\pm 1}} &= \tfrac{v_F}{2}, \\
      (A_y)_{a_m,\, b_{m\pm 1}} &= \pm\tfrac{v_F}{2}, &
      (A_y)_{b_m,\, a_{m\pm 1}} &= \mp\tfrac{v_F}{2}.
  \end{align}
  \end{subequations}
  The maximum characteristic speed is $v_F$ in every direction, which sets the Courant-Friedrichs-Lewy (CFL) bound
  on the maximum stable time step $\dd{t}$~\cite{levequeFiniteVolumeMethods2002}.

The collision term~\eqref{eq:collisions} acts diagonally on the harmonic
coefficients with the rates $\gamma_m$ defined in the main text in \eqref{eq:collisions}.

For steady-state problems, one expects that the hydrodynamic or diffusive limit requires relatively small $M$ due to fast relaxation of higher harmonics.  Still, the Fourier series coefficients for the kinetic boundary condition \eqref{eq:continuumBC} decay only as $c_m \sim 1/m$. In the strong scattering regime where $\max \gamma_m \sim 800$, we use $M = 10$ to ensure quantitative resolution of the boundary layer. In the ballistic limit $\max \gamma_m \sim 0.08$, we use $M=100$.  We interpolate logarithmically based on $\max \gamma_m$ in intermediate cases.  The potential to use relatively low $M$ is the main numerical advantage of our spectral method.

On the other hand, for the tomographic regime where $\gamma_3 < \gammaee$, we have used a constant $M=80$ to ensure the long-lived odd harmonics are captured.

\subsection{Spatial discretization}

The system~\eqref{eq:harmonic_system} is discretized in space using the
nodal discontinuous Galerkin (DG) framework provided by
Trixi.jl~\cite{schlottkelakemper2025trixi, schlottke-lakemperPurelyHyperbolicDiscontinuous2021, ranochaAdaptiveNumericalSimulations2022}.  The domain is meshed with quadrilateral
elements and within each element the solution is
represented by a degree-$p$ polynomial on Gauss-Legendre-Lobatto nodes.
Interior fluxes use the local Lax-Friedrichs (Rusanov) scheme, namely, on a face with
unit normal $\vu{n}$ pointing from a ``left'' state $\vb u_L$ to a ``right'' state $\vb u_R$, the numerical flux is
\begin{gather}
    \vb{F}^* = \tfrac{1}{2}A_n(\vb{u}_L + \vb{u}_R)
             - \tfrac{v_F}{2}(\vb{u}_R - \vb{u}_L),
\end{gather}
where $A_n \equiv n_x A_x + n_y A_y$. The collision terms \eqref{eq:collisions} are evaluated pointwise at each
quadrature node.  Time integration uses an explicit strong-stability-preserving
Runge-Kutta method; for the steady-state problems studied here, the simulation
is advanced until the $L^\infty$ residual drops below a prescribed tolerance.

\subsection{Boundary conditions}

The Boltzmann equation is a first-order hyperbolic PDE, so a boundary condition
must specify only the degrees of freedom entering the computational domain~\cite{levequeFiniteVolumeMethods2002}.  In
the continuum kinetic theory these degrees of freedom are the angular components
of the distribution with
\begin{align}\label{eq:continuumBC}
    \vu n\cdot \vu p(\theta) < 0 ,
\end{align}
where $\vu n$ is the outward unit normal to the boundary.  A harmonic
truncation does not retain this continuum half-space exactly; still, the truncated system \eqref{eq:harmonic_system} forms a \emph{discrete} hyperbolic system, where incoming and outgoing ``characteristics'' may be defined. The well-posed numerical boundary conditions---which, as $M\rightarrow \infty$, converge to the continuum result \eqref{eq:continuumBC}---must reflect the discrete characteristic structure.

At a face with outward normal $\vu n=(n_x,n_y)$, the normal streaming matrix is
\begin{align}
    A_n = n_x A_x + n_y A_y
\end{align}
where $A_x$ and $A_y$ are the harmonic-space streaming matrices appearing in
the semidiscrete kinetic equation.  We diagonalize
\begin{align}
    A_n R = R \Lambda,
\end{align}
where $R$ is the orthogonal matrix of eigenvectors and $\Lambda\equiv \mathrm{diag}(\lambda_\alpha)$ is diagonal, and we write the interior state near a given boundary in characteristic variables as
\begin{align}
    \vb w_\text{int} = R^{-1}\vb u_\text{int}.
\end{align}
The sign of each eigenvalue $\lambda_\alpha$ determines whether the
corresponding discrete characteristic is outgoing or incoming relative to the
domain.  With the convention that $\vu n$ points outward, modes with
$\lambda_\alpha>0$ leave the domain and are copied from the interior, while modes
with $\lambda_\alpha<0$ enter the domain and must be supplied by the boundary
condition.  This gives the projectors
\begin{align}
    P_+ &= R \Pi_+ R^{-1},&
    P_- &= R \Pi_- R^{-1},
\end{align}
where $\Pi_\pm$ are diagonal masks onto the eigenspaces with
$\pm\lambda_\alpha>0$.  The boundary state supplied to the numerical flux is
then constructed as
\begin{align}\label{eq:characteristic_bc_state}
    \vb u^* = P_+ \vb u_\text{int} + P_- \vb u_\text{bc}.
\end{align}
Here $\vb u_\text{bc}$ is a boundary-condition-specific target state, depending on the physics of the wall.  We have implemented several types of target state $\vb u_{\text{bc}}$, described below.

\subsubsection{Diffuse and specular walls}\label{sec:diffuseandspecular}

We model solid walls by a Maxwell boundary condition with diffuse-scattering
probability $\pscatter\in[0,1]$.  Thus $\pscatter$ is the probability for diffuse
reemission and $1-\pscatter$ is the probability for specular reflection.  In the discrete
characteristic implementation, this physical rule enters through the target
state $\vb u_\text{bc}$ in Eq.~\eqref{eq:characteristic_bc_state}.

The diffuse component is an isotropic reemission state,
\begin{align}
    \vb u_\text{diff} = c_\text{diff}\,\vb e_0 ,
\end{align}
where $\vb e_0=(1,0,\ldots,0)^T$ denotes the isotropic harmonic.  The amplitude
$c_\text{diff}$ is chosen so that the wall carries no net normal particle flux.
Since the normal current is represented in the truncated problem by the linear
functional
\begin{align}
    J_n(\vb u) = \vb q_n^T \vb u ,
\end{align}
with $\vb q_n$ the harmonic-space normal-current vector, flux balance gives
\begin{align}\label{eq:diffusewall_discrete}
    J_n\!\left(P_+\vb u_\text{int}
    + P_- c_\text{diff}\vb e_0\right)=0 .
\end{align}
Equivalently,
\begin{align}\label{eq:diffuse_amplitude_discrete}
    c_\text{diff}
    =
    -\frac{\vb q_n^T P_+\vb u_\text{int}}
           {\vb q_n^T P_-\vb e_0} .
\end{align}
This is the discrete-eigenspace version of balancing the incoming diffuse flux
against the outgoing flux.  It enforces impermeability in the same truncated
space used by the solver.

The specular component is represented by the harmonic-space reflection operator
$\mathcal R_n$.  This operator maps a distribution to its mirror image under
reflection about the wall tangent.  If $\theta_n$ is the angle of the outward
normal, the continuum action is
\begin{align}
    \theta \mapsto \theta^\text{spec}
    = \left(2\theta_n+\pi-\theta\right)\bmod 2\pi .
\end{align}
In the harmonic basis this map is a finite-dimensional linear operator acting on
the retained modes.  The specular target is therefore
\begin{align}
    \vb u_\text{spec} = \mathcal R_n \vb u_\text{int}.
\end{align}
The Maxwell wall target is
\begin{align}\label{eq:maxwell_target_discrete}
    \vb u_\text{bc}
    =
    \pscatter\,\vb u_\text{diff}
    +(1-\pscatter)\,\vb u_\text{spec}.
\end{align}
The ghost state used in the numerical flux is obtained by inserting
Eq.~\eqref{eq:maxwell_target_discrete} into
Eq.~\eqref{eq:characteristic_bc_state}.  Thus diffuse and specular scattering
only modify incoming characteristics; outgoing characteristics remain those of
the interior solution.

\subsubsection{Ohmic contact}

An ohmic contact injects carriers at a prescribed electrochemical potential
$V$.  It is implemented with the same incoming-eigenspace projection as a wall,
but with a contact target rather than a zero-flux diffuse target.  For a fully
absorbing contact,
\begin{align}
    \vb u_\text{bc} = V\vb e_0 ,
\end{align}
and the boundary state is
\begin{align}
    \vb u^* = P_+\vb u_\text{int} + P_- V\vb e_0 .
\end{align}
This corresponds to an ideal reservoir that absorbs all outgoing information and
reemits an isotropic distribution fixed by the applied bias.  A partially
absorbing contact can be represented by an absorption parameter
$p_\text{abs}\in[0,1]$,
\begin{align}
    \vb u_\text{bc}
    =
    p_\text{abs} V\vb e_0
    +(1-p_\text{abs})\mathcal R_n\vb u_\text{int}.
\end{align}
The limit $p_\text{abs}=1$ gives a perfectly absorbing ohmic contact, while
$p_\text{abs}<1$ retains a specular component at the interface.

\subsubsection{Floating voltage probe and fixed-current boundaries}

A floating voltage probe is an isolated metallic contact whose potential adjusts
so that the net current through the contact vanishes.  We implement it as an
ohmic contact with an unknown voltage $V_\text{probe}$.  After each time step, $V_\text{probe}$ is updated by solving
\begin{align}\label{eq:probe}
    \int_{\text{probe}} \dd{s}\,
    \vb J(\vb x)\cdot \vu n = 0 ,
\end{align}
using bisection.  The resulting value is the open-circuit voltage measured by a
high-impedance voltmeter.  As for the fixed-voltage contact, only the incoming
discrete characteristics are overwritten by the probe target.  A desired target current may similarly be imposed by simply adjusting the right-hand side of \eqref{eq:probe} to nonzero values.

\section{Statistical uncertainties and fitting}\label{app:inference}
In this appendix, we give the details of the noise model and fitting objective used to characterize our methods in the main text.

\subsection{Local rate uncertainties}
To determine the precision available at a fixed pair of rates, we assume independent relative standard errors $s$ in the four raw drain currents. Writing $r_a=I_a/I_N$ for $a\in\{NE,E,SE\}$, an error in the denominator $I_N$ changes all three ratios together. To leading order in $s$, their covariance is therefore
\begin{equation}\label{eq:ratio-covariance}
    (C_r)_{ab}=s^2 r_a r_b(\delta_{ab}+1).
\end{equation}
We propagate this measurement covariance through the local transport map to estimate the uncertainties in the two logarithmic rates. For $\vb x=(\log_{10}\gammamr,\log_{10}\gammaee)$, with rates expressed in units of $v_F/L$, the Jacobian is $J=\partial\vb r/\partial\vb x$. Where $J$ has full column rank, linear error propagation gives
\begin{equation}\label{eq:rate-covariance}
    C_x=(J^T C_r^{-1}J)^{-1}.
\end{equation}
The standard error of each logarithmic rate is the square root of the corresponding diagonal entry of $C_x$, in decades. Figure~\ref{fig:fractions}(e) plots their quadrature sum, $\sigma_{\mathrm{tot}}=\sqrt{\operatorname{tr}C_x}$, for $s=0.01$. This is a local statistical estimate within the assumed kinetic model. It does not include systematic errors from uncertain geometry, contact properties, wall scattering, or other model inaccuracies.

\subsection{Temperature-sweep fits}
To combine the information from different temperatures, we fit the three coefficients in the rate laws of Eq.~\eqref{eq:mock-rate-laws} jointly. We denote the dimensionless coefficients of the quadratic electron-electron term, the constant impurity term, and the linear phonon term by $\vb A=(A_{\mathrm{ee}},A_{\mathrm{imp}},A_{\mathrm{ph}})$, respectively, with rates expressed in units of $v_F/L$.

For each trial set of coefficients, the kinetic model predicts ratios $r_a^{\mathrm{model}}(T_i;\vb A)$ at the $14$ sampled temperatures $T_i=20\,\mathrm{K}+(i-1)10\,\mathrm{K}$ for $i=1,\ldots,14$. We compare their natural logarithms with the noisy observations through the residuals $\Delta_{ia}(\vb A)=\log r_a^{\mathrm{model}}(T_i;\vb A)-\log r_{ia}^{\mathrm{obs}}$. To leading order in $s$, Eq.~\eqref{eq:ratio-covariance} gives $\operatorname{Cov}(\log r_{ia},\log r_{ib})=s^2(\delta_{ab}+1)$. Using the inverse of this covariance, and taking errors at different temperatures to be independent, we minimize
\begin{equation}\label{eq:temperature-fit}
    \chi^2(\vb A)=\frac{1}{s^2}\sum_{i=1}^{14}
    \left[\sum_a\Delta_{ia}^2
    -\frac{1}{4}\left(\sum_a\Delta_{ia}\right)^2\right],
\end{equation}
where $s=0.01$ and the sums over $a$ run over $NE,E,SE$. The second term accounts for the common error in the measured $I_N$. We constrain the three coefficients to be positive and require the resulting rate trajectories to remain within the simulated range, avoiding extrapolation of the transport map.

We repeat this fit for $10^4$ independent noise realizations and evaluate the fitted rate laws at each temperature. The shaded bands in Fig.~\ref{fig:mock-experiment} contain the central $95\%$ of these recovered rates at each temperature. They measure statistical precision for the kinetic model and temperature dependences used to generate the data and do not include systematic errors from an incorrect model.

\bibliography{references}

\end{document}